\documentclass[
    reprint,
    preprintnumbers,
    superscriptaddress,
    nofootinbib,
     amsmath,amssymb,
     aps,
     prd,
    floatfix,
    ]{revtex4-2}

    \usepackage{graphicx}
    \usepackage[caption=false]{subfig}
    \usepackage[usenames,dvipsnames]{xcolor}
    \usepackage[utf8]{inputenc}
    \usepackage{color}
    \usepackage{hyperref}
    \usepackage[normalem]{ulem}
    \usepackage{physics}
    \usepackage{multirow}
    \usepackage[capitalise]{cleveref}

    \usepackage{enumitem}
    \usepackage{comment}
    \usepackage{bm}
\hypersetup{
  colorlinks   = true, %Colours links instead of ugly boxes
  urlcolor     = blue, %Colour for external hyperlinks
  linkcolor    = blue, %Colour of internal links
  citecolor   = red %Colour of citations
}
    \allowdisplaybreaks[1]
    \graphicspath{{figures/}}

\usepackage{float}
\creflabelformat{equation}{#2\textup{#1}#3}

\newcommand{\oxford}{Astrophysics, University of Oxford, DWB, Keble Road, Oxford OX1 3RH, United Kingdom}
\newcommand{\iap}{CNRS \& Sorbonne Universit\'e, Institut d’Astrophysique de Paris (IAP), UMR 7095, 98 bis bd Arago, F-75014 Paris, France}
\newcommand{\icg}{Institute of Cosmology $\&$ Gravitation, University of Portsmouth, Portsmouth, UK}

\begin{document}

\title{Optimal inflationary potentials}

\author{Tom\'{a}s Sousa}
\email{tomasfsousa@hotmail.com}
\affiliation{\oxford}
\author{Deaglan J. Bartlett}
\affiliation{\iap}
\author{Harry Desmond}
\affiliation{\icg}
\author{Pedro G. Ferreira}
\affiliation{\oxford}

\begin{abstract}
Inflation is a highly favoured theory for the early Universe.
It is compatible with current observations of the cosmic microwave background and large scale structure and is a driver in the quest to detect primordial gravitational waves. It is also, given the current quality of the data, highly under-determined with a large number of candidate implementations. We use a new method in symbolic regression to
generate all possible simple scalar field potentials for one of two possible basis sets of operators. Treating these as single-field, slow-roll inflationary models we then score them with an information-theoretic metric (``minimum description length'') that quantifies their efficiency in compressing the information in current data. We explore two possible priors on the parameter space of potentials, one related to the functions' structural complexity and one that uses a Katz back-off language model to
prefer functions that may be theoretically motivated.
This enables us to identify the inflaton potentials that optimally balance simplicity with accuracy at explaining current data, which may subsequently find theoretical motivation. Our exploratory study opens the door to extraction of fundamental physics directly from data, and may be augmented with more refined theoretical priors in the quest for a complete understanding of the early Universe.
\end{abstract}

\maketitle

\section{Introduction}
The theory of inflation has become the generally accepted explanation for the dynamics of the very early Universe \cite{Guth:1980zm,Albrecht:1982wi,Linde:1981mu}. Inflation posits that the Universe underwent an early period of accelerated expansion, driving space-time toward a flat, almost homogeneous metric. During that expansion, quantum fluctuations in the matter and the metric were stretched and amplified to macroscopic scales to become the seeds of the large scale structure we see today.

The observational evidence for inflation is compelling, albeit not definitive \cite{Planck:2018vyg}. We observe a flat, quasi-homogeneous Universe. The analysis of measurements of the cosmic microwave background (CMB) and surveys of galaxies, through various methods, indicate that the statistics of large scale structure are very much in line with what we would expect from inflation: their two point statistics are consistent with near (but not exact) scale invariance on the largest scales, and their distribution is consistent with Gaussianity. The race is now on to detect the primordial gravitational background generated during inflation. We note, however, that despite inflation's primacy in the cosmological canon, there are alternative proposals for the early Universe: the Ekypyrotic universe \cite{Khoury:2001wf}, bouncing universes (more generally) \cite{Brandenberger:2016vhg} and string gas dynamics \cite{Brandenberger:1988aj} are a few such examples.

A key aspect of the theory of inflation is its underdetermination. We have (and will have) a limited amount of information---in effect just a handful of numbers---about the properties of the Universe at early times (dark energy suffers from an analogous problem \cite{Wolf:2023uno}). Yet there has been, over the past forty years, a vast industry of inflationary model building \cite{Martin:2013tda}.
Appealing to different theoretical motivations---from string theory and supersymmetry to scale invariance, higher dimensions to modifications of general relativity, quantum gravity to grand unified theories---the list of possible inflationary theories is huge. Many of them are consistent with existing observations and will remain consistent with future observations, whatever they may turn out to be.

Before we have a working UV-complete theory, the inflationary model that one considers optimal is to some extent a matter of personal preference, depending on the theory one favours and the desiderata one has for the inflaton potential. Nevertheless it seems clear that optimality should combine in some way the notions of simplicity and accuracy---adapting Occam, the best potential should be one that is no more complex than is warranted by the data. Indeed, simplicity has been an important driver in developing theories of the early universe, although its formulation is varied. For example, in \cite{Boyle:2005ug} it was argued that the simplest models of inflation (defined in terms of the number of inflection points in the potential energy of the inflation) should have a large amplitude for the gravitational wave background; if a small amplitude (or upper bound) of the background was found, that would, according to those authors, be an indication that the theory was more complex and (more controversially) unlikely.

The goal of this paper is to quantify accuracy and simplicity self-consistently and precisely in the context of inflation, allowing a quasi-objective determination of the ``best'' potentials given current data. We achieve this by exploring
all possible single field, slow roll models of inflation constructed from particular operator sets, characterising them in terms of a precise information-theoretic definition of complexity and then ranking them. We will use a particular type of machine learning approach, {\it Symbolic Regression} (SR), to generate and test the models.
Our focus will be on studying, in an automated and systematic way, the simplest and most accurate
models of inflation and trying to determine if the sparse data that we already have are pointing us to a particular form of the theory. More rigorously, we identify the inflaton potential functions that optimally compress the information contained in current data, so that the fewest bits of information are needed to transmit the data with the help of the inflation model. This maximal data compression is the specific meaning of ``optimal'' that we employ here.

We use a particular approach for SR that we have developed, dubbed {\it Exhaustive Symbolic Regression (ESR)} \cite{Bartlett:2022kyi}. Given a basis set of operators, ESR exhaustively explores all possible functions up to a user-specified complexity. We assess the quality of each possible model using an information-theoretic metric, the {\it Minimum Description Length} \cite{RISSANEN1978,Rissanen_1983}, which tensions the likelihood of the data given the model against the model's complexity. This allows us to rank models by combining their simplicity and accuracy in a well-motivated way given the data at hand. We go further and include the intuition we have gained over decades of model building by training a language model \cite{Bartlett:2023gvh} on a large corpus of inflationary models that have been proposed in the past \cite{Martin:2013tda}. This enables us to generate prior probabilities for functions that reflect how similar they are to those appearing in the corpus, re-weighting the function ranking. This does not restrict the analysis to those functions, but rather favours the types and combinations of operators that are generated by underlying theoretical models.

There have been prior attempts at using machine learning to explore the inflationary regime. In \cite{Rudelius:2018yqi} the author used a neural network to explore the Taylor expansion of the inflationary potential, and was able to cover large field, small field and multiple field models in their foray.  The authors of \cite{Kamerkar:2022dfu} used SR to find analytic corrections to both $m^2\phi^2$ and to Starobinsky Inflation \cite{Starobinsky:1980te}, showing that the latter was robust to modifications and is a good fit to current data in and of themselves.

We structure the paper as follows. In \cref{sec:inflation} we briefly revise the theory of inflation, focusing on single field, slow roll models. In \cref{sec:sr} we describe symbolic regression, explaining Exhaustive Symbolic Regression, Minimum Description Length and the Katz back-off language model we use. We outline our problem-specific implementation of these concepts in \cref{sec:methods} and in \cref{sec:results} we apply our method to a few different basis sets and explore the range of models that are favoured. In \cref{sec:discussion} we discuss our results and look to the future at how such an approach might be developed.

\section{The Inflationary Potential and the slow roll approximation}
\label{sec:inflation}

In this paper we will restrict ourselves to a particular (but broad) class of inflationary models: single scalar field, slow roll inflation \cite{Baumann:2022mni}. The building block of such a theory is a scalar field, $\phi$, minimally coupled to gravity. On a homogeneous and isotropic background, we have that the metric is given by $g_{\alpha\beta}=(-1, a^2,a^2,a^2)$ where $a(t)$ is the time dependent scale factor of the Universe. The scale factor evolves according to
\begin{align}
    H^2&\equiv\left(\frac{\dot a}{a}\right)^2=\frac{1}{3M^2_{\rm Pl}}\left[\frac{1}{2}{\dot \phi}^2+V(\phi)\right], \\
    {\dot H}&=\frac{1}{2}\frac{{\dot \phi}^2}{M^2_{\rm Pl}},
\end{align}
where the $^.$ signifies derivative with regard to time, $M_{\rm Pl}$ is the reduced Planck mass and $V(\phi)$ is the inflationary potential that encapsulates all the information about an inflationary model. The homogeneous scalar field evolves according to
\begin{equation}
    {\ddot \phi}+3H{\dot \phi}+V'=0,
\end{equation}
where $'$ signifies derivative with respect to $\phi$.

If the kinetic energy of the scalar field is sufficiently small, the energy density will be dominated by the scalar field potential. In this {\it slow-roll} limit, we have that
\begin{equation}
    H^2\simeq \frac{V(\phi)}{3M^2_{\rm Pl}},
\end{equation}
and the Hubble rate, $H$, is almost constant; we have then that ${\ddot a}>0$.
The scalar field evolution becomes
\begin{equation}
    3H{\dot \phi}\simeq -V'.
\end{equation}
The degree of slow roll is quantified in terms of the slow roll parameters
\begin{align}
    \epsilon&=\frac{1}{2} \frac{{\dot \phi}^2}{M^2_{\rm Pl}H^2}
    \simeq \frac{M^2_{\rm Pl}}{2}\left(\frac{V'}{V}\right)^2, \\
    \eta&=M^2_{\rm Pl}\frac{V''}{V}.
\end{align}

 We define the end of inflation to be the time, $t_{\rm E}$, (or scalar field value, $\phi_{\rm E}$) at which the slow roll regime ends, i.e.
\begin{equation}
    \epsilon(t_{\rm E})=\epsilon(\phi_{\rm E})\simeq 1.
\end{equation}
The number of $e$-foldings, $N=\log(a_{\rm E}/a_{\rm I})$, between some initial (``I'') and final (``E'') time is related to the scalar field potential through
\begin{equation}
    N=\int_{\phi_{\rm I}}^{\phi_{\rm E}}\frac{d\phi}{M_{\rm Pl}} \frac{1}{\sqrt{2\epsilon(\phi)}}.
    \label{eq:efold}
\end{equation}

The Universe is not perfectly homogeneous and one must consider perturbations around the homogeneous solution of the scalar field, $\delta\phi$. Quantum fluctuations in the scalar field will interact with the fluctuations in the metric leading to  perturbations that, due to the dynamics of the background, are amplified and stretched to cosmological scales. One can characterise the power spectrum of the fluctuation in the scalar field, on superhorizon scales and at the end of inflation,
\begin{equation}
    \left(\frac{H}{\dot \phi}\right)^2\langle|\delta\phi(k)|^2\rangle \equiv A_S \left(\frac{k}{k_*}\right)^{n_S-1},
\end{equation}
where we have taken the Fourier transform of $\delta\phi$, the pivot scale is $k_*=0.05$ Mpc$^{-1}$ and we have defined the amplitude, $A_S$, and the scalar spectral index, $n_S$, of the perturbations. These are related to the slow roll parameters via
\begin{align}
    A_S&= \frac{1}{8\pi^2}\frac{1}{\epsilon}\frac{H^2}{M_{\rm Pl}} \label{eq:A_Stheory}, \\
    n_S-1&=-6\epsilon\left(\phi_{I}\right) + 2\eta\left(\phi_{I}\right) \label{eq:n_Stheory},
\end{align}
where all the quantities are evaluated at the time, $t_I$, during inflation, that a given scale exits the horizon.

Finally, one finds that tensor fluctuations are also excited during inflation leading to a bath of primordial gravitational waves also characterised by an amplitude given by
\begin{equation}
    A_T=\frac{2}{\pi^2}\frac{H^2}{M^2_{\rm Pl}}. \label{eq:A_Ttheory}
\end{equation}
It is useful to define the relative amplitude of tensor to scalar fluctuations in terms of
\begin{equation}
    r\equiv\frac{A_T}{A_S}=16\epsilon\left(\phi_{I}\right). \label{eq:rtheory}
\end{equation}
From now on, we will rescale $\phi\rightarrow \phi/M_{\rm Pl}$ and, in doing so, remove $M_{\rm pl}$ from all the expressions. We then have that scalar field variations will be in units of $M_{\rm Pl}$.

Over the past few years, it has been possible to place constraints on some of these parameters with high-resolution, high-precision measurements of the anisotropies in the cosmic microwave background (CMB) radiation \cite{Planck:2018vyg} combined with gravitational wave experiment constraints. We currently have that \cite{Planck_2018_Inflation, Galloni_2023}
\begin{align}
    A_S &= \left(0.027 \pm 0.0027\right) M_{\rm Pl}, \label{eq:A_S} \\
    n_S &=0.9649 \pm 0.0042, \label{eq:n_S}  \\
    r &<0.028\ (95\%\,\text{CL}).  \label{eq:r}
\end{align}
We will use these constraints (which we refer to as ``the data'' hereafter) to determine the optimal functional form of the inflationary potential.

\section{Symbolic Regression}
\label{sec:sr}
Machine learning with SR attempts to infer mathematical formulae from data
\cite{Petersen_2021,Landajuela_2022,Tenachi_2023,Biggio_2021,Worm, Kammerer_2021, Dome,FFX,aifeyn_0,aifeyn,QLattice,Lemos_2022,cranmer2020discovering,pysr,Cranmer_2023,DM,Schmidt_2009,Schmidt_2011,Virgolin_2019,deFranca_2021,LaCava_2019,Kommenda_2020,Virgolin_2021,Arnaldo_2014,Bartlett:2022kyi}.
The aim is to determine the explicit mathematical expressions that optimally combine simplicity and accuracy on a dataset and which, given their simplicity, might shed light on the processes that led to the emergence of the data. The approach differs from other methods in that the final expression is not pinned down (or restricted) to a linear combination of a restricted set of basis functions (that one might get in, say, principal component analysis) or the composite combination of simple basis functions (as one might find in neural networks).

In SR, one begins by choosing a basis set of operators, such as, for example, $\{\phi,c,*,+,/,-,\exp,\log, {\rm power}\}$ (where $c$ is a constant) and then constructs
possible combinations of these operators of a given complexity, which we define as the number of operators (including variables and parameters) that the expression contains. So, for example, $V(\phi)=\phi$ or $V(\phi)=c$ have complexity 1, $V(\phi)=\exp(\phi)$ has complexity 2, $V(\phi)=\phi+c$ is complexity 3, etc.  It is instructive to look at a few well known potentials. With this basis, {\it quadratic}, $V(\phi)\sim m^2\phi^2$ or {\it quartic}, $V(\phi)\sim \lambda\phi^4$ inflation potentials have complexity 5, {\it power law} inflation potential, $V(\phi)\simeq Ae^{-\lambda \phi}$ has complexity 6 while {\it Starobinsky inflation} (or {\it Higgs} inflation \cite{Bezrukov:2008ej}), $V(\phi)\simeq A [1-\exp(-\sqrt{2/3}\phi)]^2$ has complexity 10.  We note that  {\it $\alpha$-attractor inflation} \cite{Kallosh:2013yoa}, while possessing several attractive properties, cannot be classified in this way as it has been argued that its predictions are, effectively, independent of the shape of the potential \cite{Kallosh:2021mnu}. We are therefore not able to show quantitative results for it.

A few comments are in order about this definition of complexity. First, it characterises the structure of the function and not its behaviour: structurally simple functions with few operators may produce ``complex'' behaviour such as many inflection points or a high degree of nonlinearity. (In the language of genetic programming, our complexity definition reflects the {\it genotype} of the function and not its {\it phenotype}.)
Second, the complexity of the expression depends on the set of basis functions with which it is represented. For example, $V(\phi)=\tanh(\phi)$ has complexity 2  if $\tanh$ is included in the basis set but complexity 13 if it must be built up from $\exp(\phi)$ and other elementary operators.

Classifying potentials in terms of their complexity can be done systematically but, as one might imagine, the higher the complexity, the wider the range of expressions that must be considered. When should one stop searching for the best mathematical expression given the data? For a fixed complexity, there is always an expression that maximises the likelihood of the data, ${\cal L}(D)$. The set of such functions over a range of complexities
defines a line in the complexity-likelihood plane, which is called the {\it Pareto front}. Functions that lie on that line are called {\it Pareto optimal}, and \textit{Pareto dominate} all the worse functions that lie at lower likelihood values.

Different methods have been used to search for the Pareto front, including supervised or reinforcement learning with neural networks \cite{Petersen_2021,Landajuela_2022,Tenachi_2023,Biggio_2021}, deterministic approaches \cite{Worm, Kammerer_2021, Dome,FFX}, Markov chain Monte Carlo \cite{Jin_2019}, and physics-inspired searches \cite{aifeyn_0,aifeyn,QLattice}.
 The most popular approach \cite[e.g.][]{Lemos_2022,cranmer2020discovering,pysr,Cranmer_2023,DM,Schmidt_2009,Schmidt_2011,Virgolin_2019,deFranca_2021,LaCava_2019,Kommenda_2020,Virgolin_2021,Arnaldo_2014} involves genetic programming \cite{turing,David, haupt} which, roughly speaking, involves breeding and mutating expressions until one or more of sufficiently high fitness (likelihood) have been produced.
 Genetic  programming can be used to scan a vast number of expressions but suffers from some major problems. To begin with, it has a propensity to search for ever more complex expressions, a problem known as ``bloat.'' As a consequence of this property, it does not thoroughly and exhaustively search all the expressions at the lowest complexity. In addition, genetic programming offers no guarantee of finding the Pareto optimal function at any given complexity and so the robustness of the results is very difficult to establish.

In an attempt to mitigate the issues facing genetic programming, we have developed a new approach which can a) scrutinise {\it all} functions at given complexity, b) rigorously assess the trade off between how well they fit the data and how complex they are, and c) incorporate prior assumptions about how reasonable an expression is, given the physical context one is exploring. We now briefly describe each of these aspects in turn.

\subsection{Exhaustive Symbolic Regression}
\label{sec:esr}
With Exhaustive Symbolic Regression (ESR) \cite{Bartlett:2022kyi}, one systematically constructs all possible functions of a given complexity and calculates their likelihood to be sure of finding the Pareto optimal function at that complexity. In practice one generates all possible functions with a given number of operators and then eliminates duplicates and functions that can be simplified.

There are a few points that we should mention. First of all, one needs to choose a basis set of operators from which one can build the space of expressions that one wishes to explore. In other words, there is a {\it prior} dependence that must be taken into account in such an analysis. This can be mitigated by considering different possible sets and checking what the Pareto front looks like for each choice.

Furthermore, while the process can be systematised and made efficient, due to the fact that the number of possible functions grows exponentially with complexity there is a limit to how high a complexity one can go. This occurs simply for computational reasons, and in our case the limit is at complexity $\sim$8--9.
This may not be a serious concern. Looking at a plethora of different expressions arising in a range of physical problems, one typically finds that they are of low enough complexity to be within the scope of ESR. Nevertheless, it is important to bear this limitation in mind in any ESR analysis.

\subsection{Minimum Description Length}
\label{sec:mdl}
The goal is to determine the optimal inflationary potential, given the current data. To achieve this, we need to define how to trade-off a function's likelihood, or accuracy at accounting for the data, with its complexity. We have proposed the \emph{Minimum Description Length (MDL)} principle for this purpose \cite{RISSANEN1978,Rissanen_1983,Bartlett:2022kyi,Bartlett:2023gvh}. MDL has an information theoretic motivation and gives a model selection criterion which makes commensurable the two objectives of maximising accuracy and simplicity. MDL seeks the function that minimises the \emph{description length (DL)}, which is the number of units of information needed to specify or communicate the data with the help of the function.
The DL of a function includes contributions from its structure and parameters, measuring simplicity, and the residuals of the data around the function's expression, measuring accuracy. In the simplicity terms, the use of more operators and more parameters (especially those that must be specified precisely to achieve high accuracy) is penalised.
The residuals are encoded using the Shannon--Fano coding scheme, which penalises inaccurate functions according to their negative log-likelihood.

The resulting expression for the DL, $L(D)$ is
\begin{equation}
    \label{eq:mdl}
    \begin{split}
        L(D)=&-\log({\hat {\cal L}})+k\log(n)-\frac{p}{2}\log(3) +\sum_j\log(c_j) \\
        &+\sum_i^p\left[\frac{1}{2}\log({\hat I}_{ii})+\log(|{\hat \theta}_i|)\right]
    \end{split}
\end{equation}
where $D$ is the dataset, ${\cal L}$ the likelihood, $\theta_i$ a free parameter in the expression (of which there are $p$), $k$ the number of nodes in the expression's tree representation, $n$ the number of unique operators involved, $I$ the Fisher information matrix of the parameters and $c_j$ a constant natural number generated by simplifications. Hat denotes evaluation at the maximum likelihood point. We are using natural logarithms, so the DL is the number of nats needed to communicate the data with the aid of the expression. For a derivation of \cref{eq:mdl} and further discussion, see~\cite{Bartlett:2022kyi}.

The fact that we are balancing the two aspects of an expression---its goodness of fit versus its complexity---means that the DL has an appealing property: if the complexity is too low, the expression will be a bad fit to the data and the DL will be large. If the complexity is too high, the simplicity terms will again make the DL large. For a given dataset and operator basis set one therefore expects a minimum in the DL.
Note that this is not a minimum at each separate complexity (as is the case for likelihood) but rather a \emph{global} minimum occurring at some complexity and which corresponds to the overall best function. With the Minimum Description Length principle, model selection therefore becomes a completely unambiguous procedure and one is left with the purely technical challenge of identifying the MDL function.

One can get a rough intuition of how the two parts of the DL play against each other. If the data are not particularly constraining (few data points or large uncertainties), the likelihood term will be small and hence $L(D)$ will be dominated by the complexity terms. Thus very simple functions will be favoured, as would be expected from the desideratum to prevent overfitting. On the other hand, if there is an abundance of well measured data the likelihood term will be exacting and high complexity expressions that fit the data well will be favoured. The MDL principle therefore enables us to identify the level of complexity of the function (in this case inflationary potential) that is warranted by the strength of the data. We consider inflaton potentials ``optimal'' to the extent that they have low DL on the current data.

Minimising the DL is an alternative to the more commonly used model selection method based on the Pareto front.
Increasing from low values of complexity, the Pareto front will typically show a decrease in $-\log{\cal L}$ (or e.g. the mean square error in case the data do not have proper uncertainties). For some value of the complexity, there is typically a large drop in value in $-\log{\cal L}$---a ``knee''---followed by a slowly decaying plateau. From the plateau onward, there is no great reward in increasing the complexity and so one definition of the overall best model is the one at the knee. Although not requiring computation of any of the terms in the DL besides the log-likelihood, this is an inferior model selection procedure to MDL because knee behaviour is not generic in the Pareto front and this approach has no statistical or information-theoretic justification. Thus, while we will discuss potentials at the Pareto knees, we stress that the MDL potentials should be considered superior.

\subsection{Katz Model}
\label{sec:katz}
With symbolic regression, it is possible to generate a huge variety of functional forms, many of which bear little resemblance to functions likely to be generated by an underlying theory.
For example, a function such as $f(x)=\cos(\theta_1\cos(\theta_2\cos(\theta_3 x)))$ (where $\theta_1$, $\theta_2$ and $\theta_3$ are parameters) is extremely unlikely to be relevant in most applied scenarios, yet is fairly simple and could be favoured by MDL. It would be useful to have a principled way to weed out such
functions from the family of expressions we are assessing.

ESR (as with almost all machine learning) is purely empirical, and it is not within scope to determine the theoretical validity or significance of every candidate function. We therefore need a proxy for physical motivation that depends purely on the functions themselves, for which we choose similarity to inflaton potentials that have already been published. To understand how this information may be used, it
helps to think of expressions as sentences or phrases in a particular language: there are word (or operator) sequences that ``make sense'' where others do not.
With this perspective in mind, we have adapted a simple language model trained on a corpus of expressions to upweight combinations and sequences of operators that are common and hence more likely to make theoretical sense. This does not mean that only expressions that are present in the training set are acceptable, but rather that the sequences of operators found there are favoured when evaluating the functional parameter space.

To be specific, we use a Katz back-off model \cite{Katz_1987,Bartlett:2023gvh} to estimate the probability of any argument to an operator conditioned on the operator's identity.
For each operator appearing in the expression, we first find its parent, grandparent etc. until we are at the ``root'' operator, i.e. the operator which is not an argument of any other function or have reached a list of length equal to the total number of operators in the function, $n$. We also include the ``sibling'' operators in this list to capture this conditional probability. In this work we choose $n=9$ given that we analyse function trees up to depth 9, but have verified that our results are not sensitive to this choice. This forms an $n$-gram of operators, whose probability can be estimated from the number of occurrences of such an $n$-gram in a corpus (or training set) of equations chosen for a given context. If the $n$-gram does not appear in the corpus, the algorithm ``backs off'' to the $(n-1)$-gram and repeats this process until the sequence has been seen. The conditional probabilities of each operator are then combined to assign a single probability to the function. This behaves as a prior on the function's probability of being correct.

We perform our analysis with and without the Katz back-off model, to study how the prior assumptions about the potential's structure will affect our results. Without the language model, we take the MDL expression from \cref{eq:mdl}.
Since the description length can be viewed as an approximation to the Bayesian evidence with a particular choice for the prior on functions \cite{Bartlett:2023gvh}, when we use the language model we replace the term giving the structural complexity of the function ($k\log n$)
by $-\log \Pi_i,$ where the $\Pi_i$ is the prior resulting from the Katz back-off model. From an information-theory perspective, this can be viewed as sending the structural information of the function with a code designed such that the \textit{a priori} most likely functions (with reference to the corpus) have the shortest code lengths. We choose as our corpus the set of inflationary potentials listed in the \textit{Encyclopaedia Inflationaris} \cite{Martin:2013tda}, as described in more detail below.

\section{Methods}
\label{sec:methods}

\begin{figure}
    {\includegraphics[width=1\linewidth]{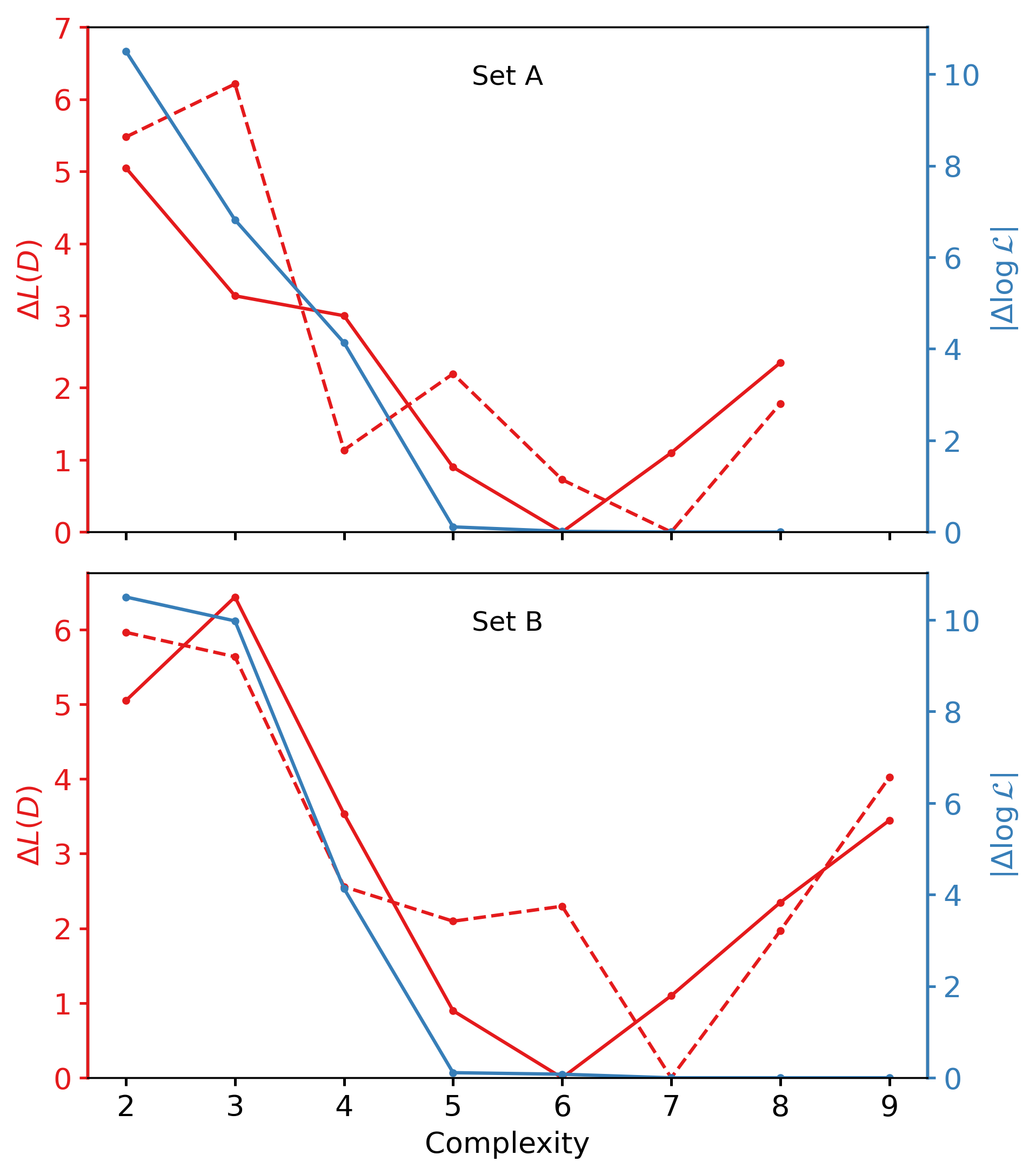}}
    \caption{Pareto front of inflationary potentials found with ESR when compared to the data for the two basis sets. We show the best-fitting functions according to the change in the description length, $L(D)$, (red) and the likelihood, $\mathcal{L}$, (blue) relative to the corresponding minima. More accurate functions appear at lower $|\Delta\log(L)|$, while overall superior functions appear at lower $L(D)$. For the solid line we use the $k\log(n)$ term in the description length to penalise model complexity, while for the dashed line we instead use the Katz language model.
    }
    \label{fig:pareto}
\end{figure}

We can now describe the ingredients of our analysis. We take the likelihood of the data, ${\cal L}$, to satisfy
\begin{equation}
    -\log {\cal L}=\frac{(n_S-n^e_S)^2}{2\sigma^2_{n}}+\frac{(r-r^e)^2}{2\sigma^2_{r}}+\frac{(A_S-A^e_S)^2}{2\sigma^2_{A}},
\end{equation}
where the measured values ($A^e_S$, $n^e_S$ and $r^e$) and uncertainties are listed in \cref{eq:A_S,eq:n_S,eq:r}
and the model parameters ($A_S$, $n_S$ and $r$) are derived from the inflationary potential through \cref{eq:A_Stheory,eq:n_Stheory,eq:A_Ttheory,eq:rtheory}.
Since the measured value of $r$ is consistent with zero, we use $r^e=0$ for this term in the likelihood, corresponding to Gaussian centred at 0.
Although the full data likelihood is neither exactly separable nor Gaussian,
it is sufficiently nearly so for this approximation not to affect our results.

The algorithm we use to compute the predicted $A_S$, $n_S$ and $r$ values from a potential works as follows. We first find all $\phi_{E}$ values that satisfy the condition $\epsilon \left(\phi \right)=1$. Then, for each of the $\phi_{E}$ we use \cref{eq:efold} to find the corresponding value of $\phi_{I}$, i.e., where observable inflation begins, by looping over a guess for $\phi_{I}$ until the threshold $N=60$ is reached. To obtain the exact value of $\phi_{I}$ we find the root of $N-60$ between the interval $\phi_{E}$ and the last guess for $\phi_{I}$. At this point we can simply compute $A_S$, $n_S$ and $r$ from \cref{eq:A_Stheory,eq:n_Stheory,eq:rtheory}. In case the potential allows for more than one inflationary trajectory, we consider the one whose predictions better match the observations.

We consider two different operator basis sets:
\begin{eqnarray}
& &\Big\{x,a,{\rm inv},\exp,\log,\nonumber \\
& &\ \ \ \sin,\sqrt{|.|},{\rm cube},{\rm square},+,*,-,/\Big\},  \  \ \ \mbox{Set A,} \nonumber \\
& &\Big\{x,a,{\rm inv},\exp,\log,+,*,-,/,{\rm power}\Big\},  \ \ \ \ \ \ \ \mbox{Set B.} \nonumber
\end{eqnarray}
There are trade-offs in the choice of each basis set. Set A permits only analytic functions for the inflationary potential, in line with what is usually seen in inflationary model building. The downside is that the basis set is large which greatly limits the maximum complexity we can reach with ESR.
In Set B we loosen this restriction by allowing
the ``power'' operator
\begin{equation}
    {\rm power}[f(x),g(x)]=\vert f(x) \vert^{g(x)},
\end{equation}
The basis set is much smaller allowing us to reach higher complexity but it does allow for more unconventional potentials. The hope, then, is that the language model can limit the excesses of such freedom and prioritise functions which make more sense, given the existing inflationary canon. We remove functions that are the same as others under $\phi \rightarrow -\phi$, since they have the same Lagrangians up to a change of variable, and flag as undesirable potentials that have negative values for the maximum-likelihood parameter values or that are unbounded in $\phi$. We do not flag potentials with other potential pathologies but which nevertheless appear in the Encyclopedia Inflationaris. The reader may make whatever further cuts on our results tables that he or she desires.

For each model, we need to find the maximum likelihood values of the parameters in the potential. We employ the BFGS optimisation algorithm \cite{B,F}, using up to $N_{\rm iter}$ different possible starting points to increase the probability of finding the global minimum for a given potential. If $N_{\rm conv}$ iterations of the optimiser give a $\log \mathcal{L}$ within 0.5
of the best solution found so far we conclude that the optimiser has found the global best-fit parameters and terminate the optimisation early.
These are, then, the parameters that go into calculating the DL which we use to classify and rank the inflationary potentials.
We choose $N_{\rm iter}= 60 p + 40$ and $N_{\rm conv} = 20p - 5$ for functions with $p$ parameters, since in general it is harder to find the global optimum in a higher-dimensional space.

For parameters that are consistent with zero within the estimated uncertainties, we check whether these can be set identically to zero since this can in principle lower the description length of the function. If the DL is reduced, then the parameter is kept to zero, otherwise we use its non-zero value.
If a parameter is consistent with zero but setting it to zero would create a potential unfit for inflation (such as when the parameter multiplies the entire expression), we redefine the parameter's uncertainty to be its value, corresponding to the knowledge that it must be
greater than zero.
This gives a fixed value of $\log 2 \approx 0.69$ nats for that parameter's codelength (corresponding to the term $p\log 2$ in Eq.~2 of \citet{Bartlett:2022kyi}).
This effect arises from a breakdown of the approximation that the posterior distribution of the parameter can be represented as a Gaussian centred at the maximum likelihood point with a covariance given by the second derivative of the likelihood at that point.

To train the Katz back-off model we use a list of theoretically motivated potentials put together in an extensive review of inflationary models \cite{Martin:2013tda}.  We express the potential functions as trees made up of the same basis operators as the functions whose prior will be calculated. This means that we remove all models with expressions such as $\phi^{\theta_0}$ (where $\theta_0$ is any constant) from the training set to be used on Set A functions, since Set A only allows for exponents which are a combination of square, cube, square root and inverse. Similarly, we do not include the models with trigonometric functions from \cite{Martin:2013tda} on the training set to be used with Set B functions. In total we have 45 models in the training set for Set A functions, and 58 models for Set B.

\section{Results}
\label{sec:results}

\subsection{The optimal inflationary potentials}

\begin{figure*}
    \centering
    \begin{tabular}{ccc}
        \includegraphics[width=0.3\textwidth]{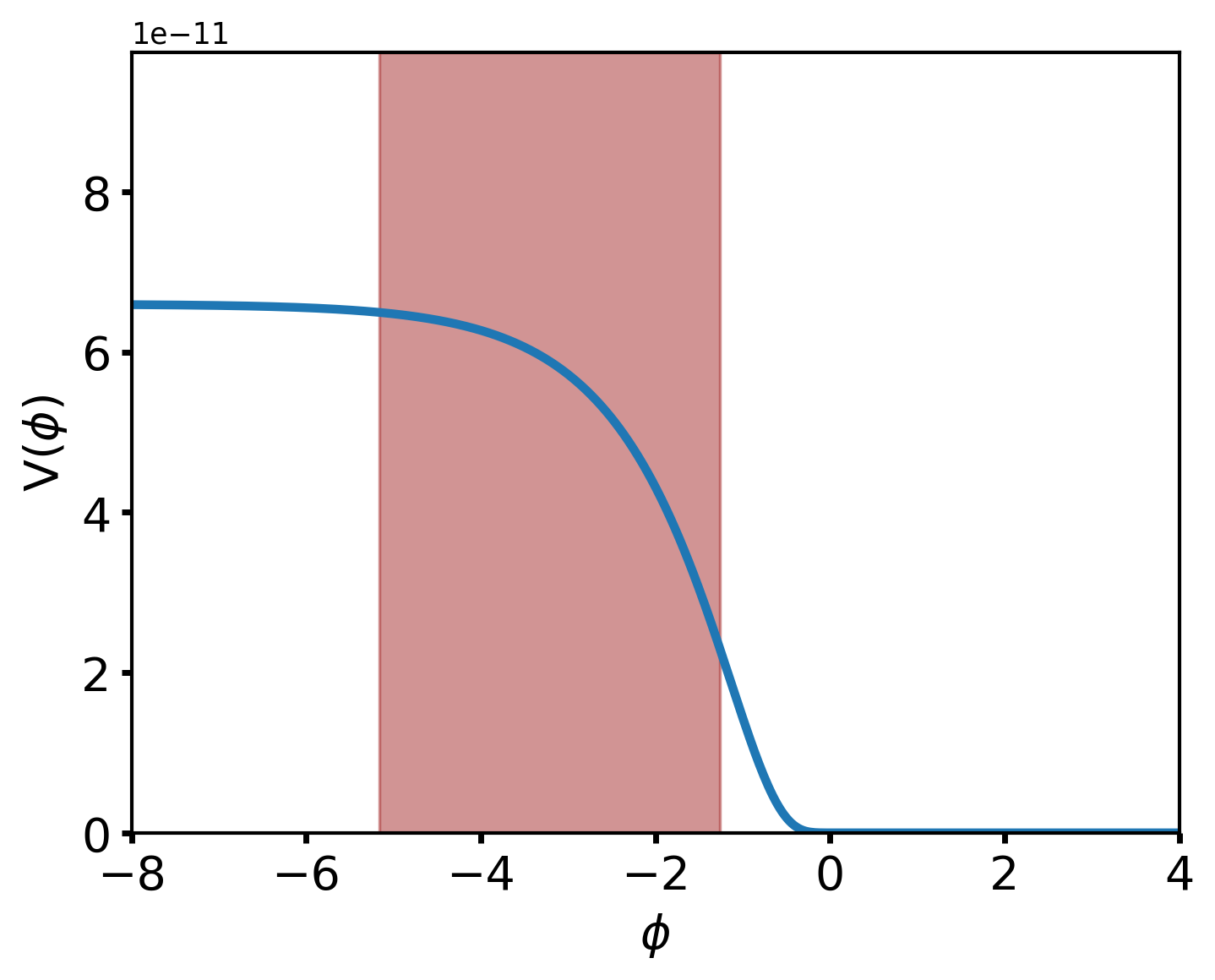} & \includegraphics[width=0.3\textwidth]{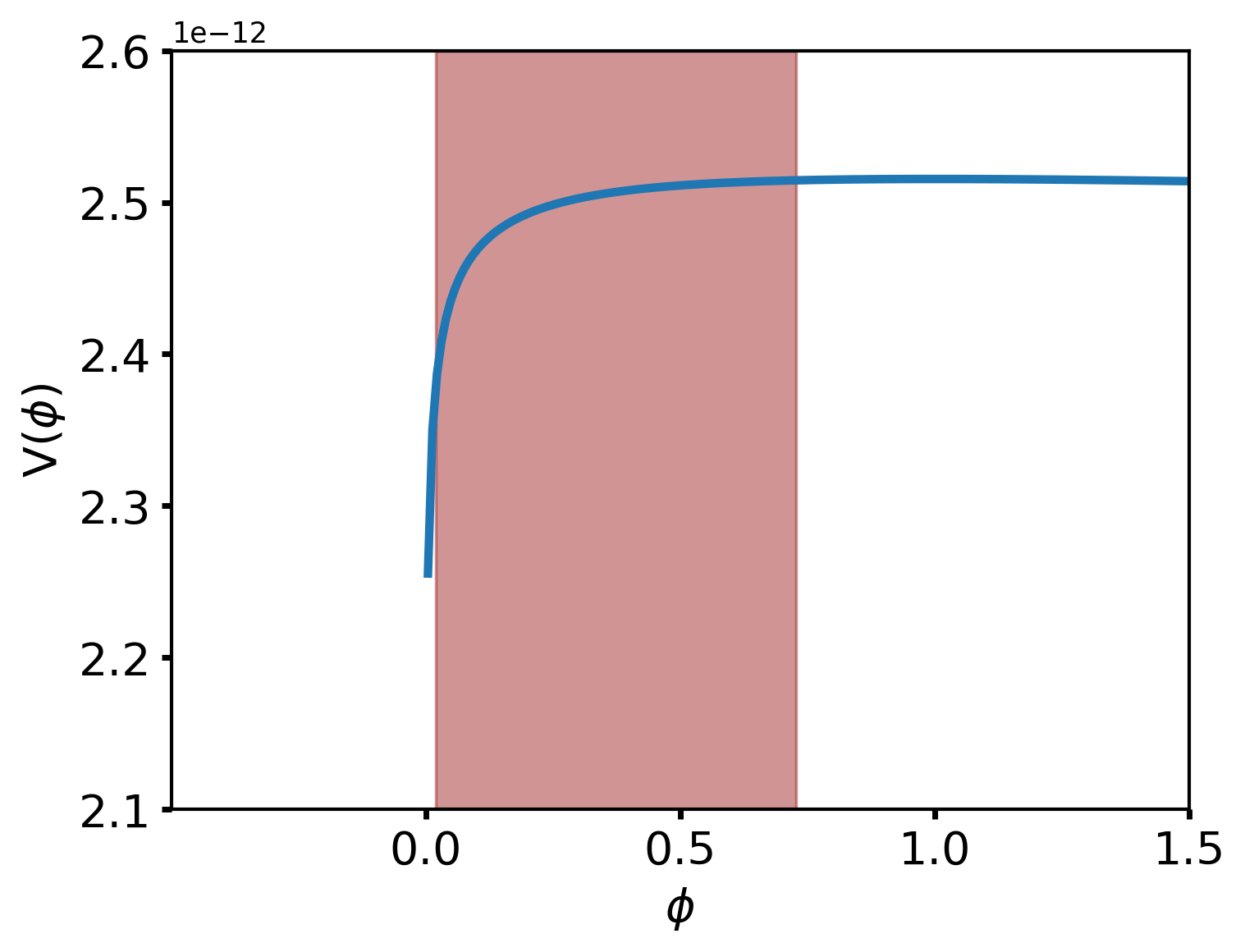} & \includegraphics[width=0.3\textwidth]{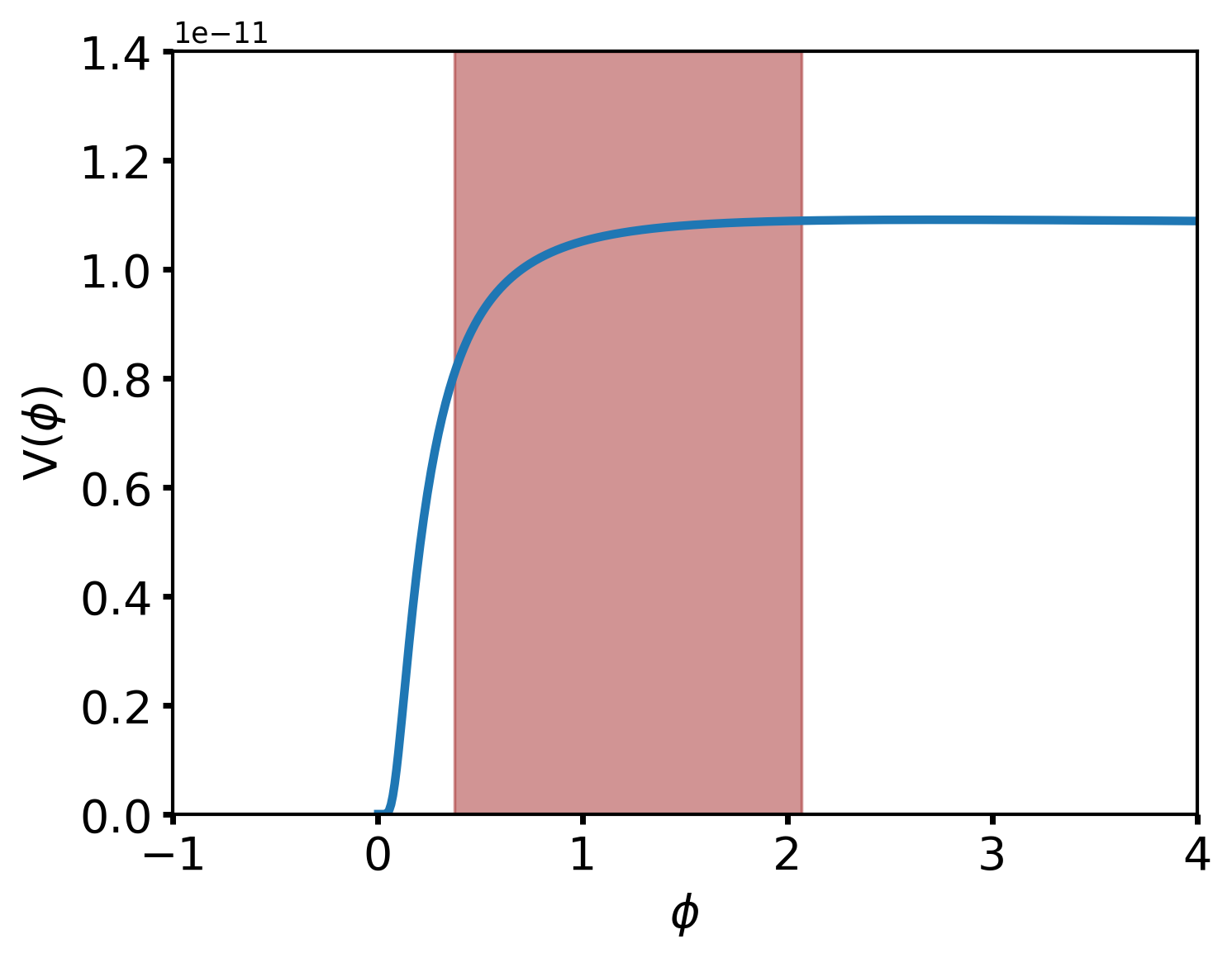} \\
        (a) $V\left(\phi \right)=e^{- e^{e^{e^{\phi}}}}$ & (b) $V\left(\phi \right)=\theta_{0} \left(\theta_{1} + \log{\left(\phi \right)}^{2}\right)$ & (c) $V\left(\phi \right)=\theta_{0} \phi^{\frac{\theta_{1}}{\phi}}$
    \end{tabular}
    \begin{tabular}{cc}
        \includegraphics[width=0.3\textwidth]{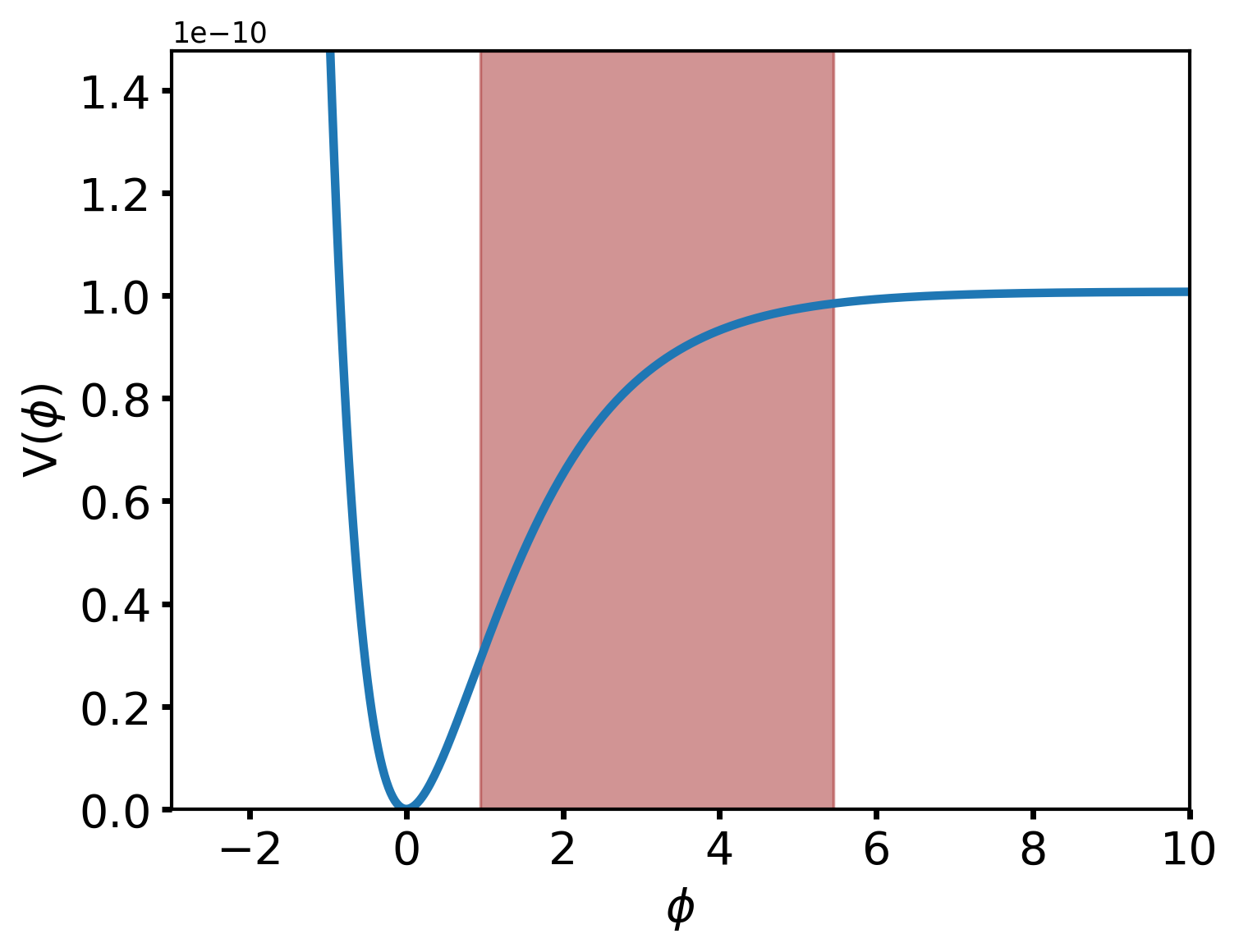} & \includegraphics[width=0.3\textwidth]{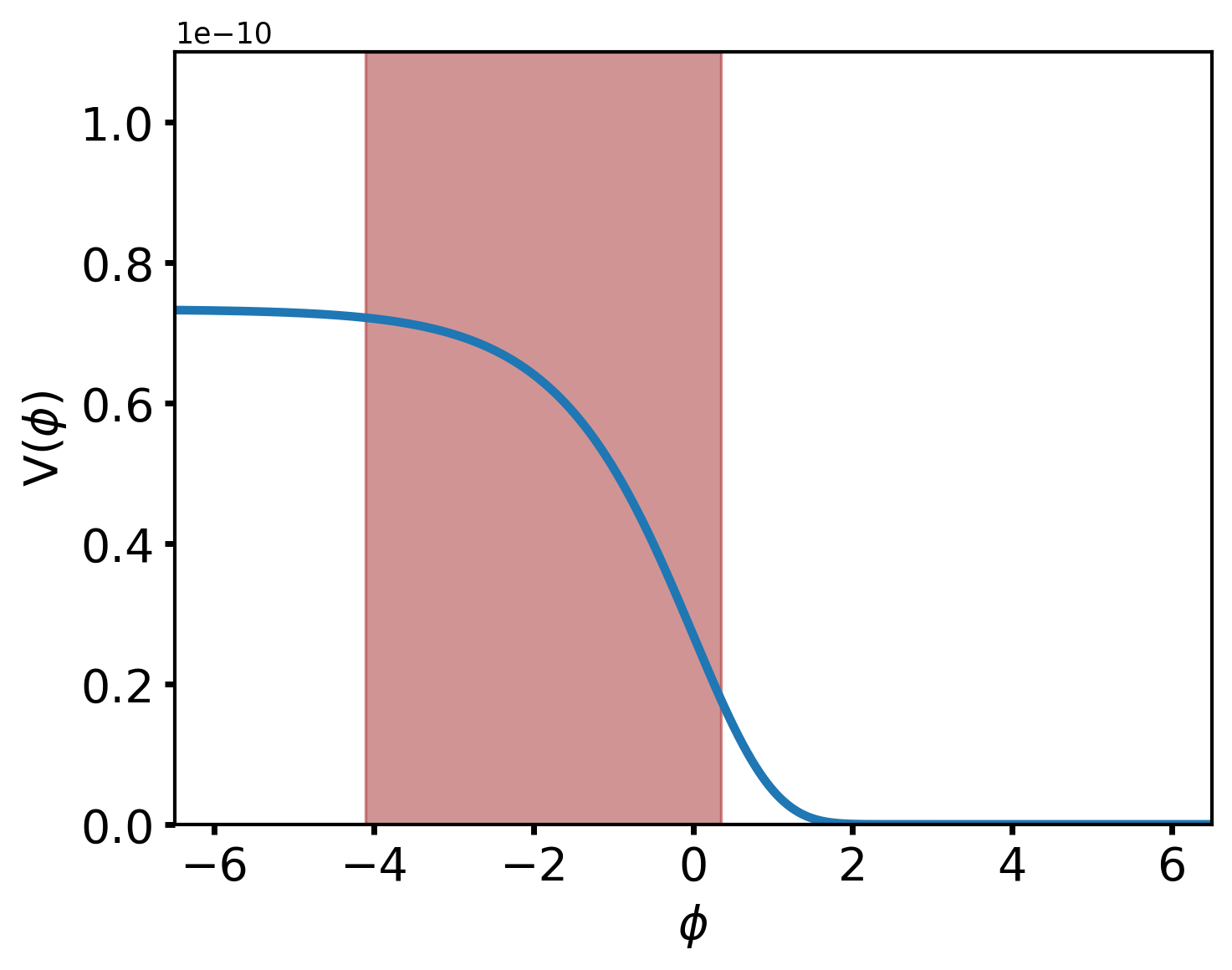} \\
        (d) $V\left(\phi \right)=\theta_{0}\left(1-e^{-\sqrt{2/3}\phi}\right)^2$ & (e) $V\left(\phi \right)=\theta_{0}e^{- e^{\phi}}$
    \end{tabular}
    \caption{Inflationary trajectories of the potentials with the Minimum Description Length for different analyses:
  (a) The MDL function for both basis sets A and B using the $k\log(n)$ function prior, (b) The MDL potential for basis set A with the language model prior, and (c) The MDL expression for basis set B with the language model. For context, in (d) we give the corresponding plot for the Starobinsky or Higgs inflation,
  with $\theta_{0}$ optimised to fit $A_S$, and in (e) we give the potential which is at the ``knee'' of the Pareto front (\cref{eq:bestfit}).
  The shaded region shows the range of $\phi$ during which inflation occurs, where the inflaton rolls from the region of high to low potential.
  }
  \label{fig:plots}
\end{figure*}

Let us now turn to the results. 
In \cref{fig:pareto} we present the Pareto front and the curves of DL vs complexity for different combinations of priors (i.e. of the basis sets $A$ and $B$, with and without language priors). 
In this section we discuss a sub-set of these potentials but tabulate the ten highest ranked functions for these choices in \cref{tab:SetA klogn,tab:SetB klogn,tab:SetA katz,tab:SetB katz} of \cref{appendix}.
Using the $k\log(n)$ rather than the language model prior, the function that minimises the DL for both operator sets occurs at complexity 6:
\begin{equation}
    V= e^{-e^{e^{e^\phi}}}. \label{eq:bestfitDLnL}
\end{equation}
This nested composition of exponentials is unlike any inflationary potential that has been proposed until now, indicating that it may not have a physical motivation. Nevertheless, in panel (a) of \cref{fig:plots} we see that it has the shape characteristic of slow roll potentials. There are no free parameters (we choose to express all potentials in units of $10^{-9}M_{\rm Pl}$ for numerical stability)
and the change in the scalar field during inflation is trans-Planckian ($\Delta \phi=|\phi_I-\phi_E|>1$), showing it to be a large-field model.
While the potential in \cref{eq:bestfitDLnL} is the optimal model for both Set A and B, we can also see this propensity for nested exponentials as we go down the ranking for each set, which are given in \cref{tab:SetA klogn,tab:SetB klogn}.
There are some differences between the sets however. In Set A, for example, we find a few potentials which bear a passing resemblance with those explored in the literature (such as $V(\phi)=\theta_0-e^{2\phi}$ or $V(\phi)=\theta_0-e^{3\phi}$), while Set B is totally dominated by nested functions.

Using the ``Pareto knee'' method of model selection would instead pick out a function at complexity 5:
\begin{equation}
    V=\theta_0 e^{-e^\phi}, \label{eq:bestfit}
\end{equation}
where $\theta_0$ is the only free parameter. This potential is a slightly less extreme version of \cref{eq:bestfitDLnL} with two fewer nested exponentials, and has best-fit parameter values giving $n_S=0.9668$ and $r=0.002$.
Although there is a very small improvement in likelihood between this function and the best potential at complexity 6, we find that the change in the description length is much larger, such that this is not the MDL expression. From \cref{tab:SetA klogn} we see that this is due to two effects. First, the additional parameter in this expression must be transmitted which adds to the codelength of the function, and thus the expression is deemed more complex. Moreover, the nested exponential function in \cref{eq:bestfitDLnL} contains fewer unique operators than \cref{eq:bestfit}, and thus is less severely punished by the $k\log n$ term in \cref{eq:mdl}.
It is instructive to modify it slightly by replacing $\phi$ by $\theta_1\phi$. This adds an extra parameter and thus, increases the total codelength: the complexity of the potential increases by $2$, and the parameter contribution to the codelength will increase the overall DL. {If one does that, one finds that $(\theta_0,\theta_1) =(0.175,0.650)$ and, while $n_S$ remains unchanged, we find that $r\simeq 5\times 10^{-3}$. However, the scalar field trajectory remains trans-Planckian. If one chooses $\theta_1\simeq 10$ one can make $\Delta\phi<1$ and avoid large field excursions; for this particular choice of $\theta_1$ one has $r\simeq 10^{-5}$.}
The desire (and ability) to control for trans-Planckian behaviour in the priors will be discussed in \cref{sec:discussion}.

We now turn to the effect of using the language model on potential selection and ranking. If we look at \cref{tab:SetA katz}, in the case of Set A, we find that language model has a marked effect: the preferred potentials are similar, but not identical to the type of potentials that make up the corpus of models on which the language model was trained. This removes the preference for potentially unphysical nested exponentials, indicating that functions' structures are likely correlated with their degree of physical motivation. We find variants of Starobinsky and Higgs inflation as well as some novel potentials involving logarithms. It is also interesting that one can find one of the most basic potentials---the double well potential \cite{Linde:1994wt}, consisting of a sum of quadratic and quartic terms---amongst the most highly ranked; individually, the quadratic and quartic potentials lead to an unacceptably high value of $r$, but combined appropriately, they seem to just scrape within what is acceptable by current data. Unsurprisingly $r$ for this model is one of the highest on the list.

Given the limitations we have in terms of complexity, there is tentative evidence that the  minimum of the DL in Set A with the Katz prior now occurs at complexity 7, corresponding to a potential of the form
\begin{equation}
    V = \theta_0\left[\theta_1+\log(\phi)^2\right].
\label{eq:bestfitDLALM}
\end{equation}
The logarithm dependence on $\phi$ is evocative of the type of potentials one might obtain from radiative corrections, such as ones that arise in Coleman-Weinberg potentials. But there, the logarithms normally arise on their own, and not raised to any power. Powers of logarithms can, however, appear in string inspired models  and were considered in the early days of supersymmetric models of inflation \cite{Witten:1981nf,Albrecht:1983ib} as well as later on \cite{Gerasimov:2000zp}. In fact it is curious to observe that, for reasonable choices of the field content in the supersymmetric models, the potentials arising in \cite{Witten:1981nf} have an almost identical structure to what we find here, even if such a potential does not explicitly appear in the corpus we have used to train the language model.
The best fit parameters are such that this potential is unbounded from below and higher order corrections would be necessary for this model to be stabilized.
As can be seen in \cref{tab:SetA katz}, $r<10^{-4}$, i.e. the lowest on the list and lower than the values that we saw in the case of the $k\log n$ prior.

For Set B, the minimum value of the DL occurs at complexity 5 with the function
\begin{equation}
    V(\phi)=\theta_{0} \phi^{\frac{\theta_{1}}{\phi}}, \label{eq:bestfitBDLLM}
\end{equation}
Note that, even with the language model, a set of nested functions, unlike those proposed in the training corpus, is still favoured by the data. We find that this model has a very low value of $r$. While the minimum DL curve is not monotonic, we are confident that we have found the optimal model for this set of priors.

For reference, in \cref{tab:SetA klogn,tab:SetB klogn,tab:SetA katz,tab:SetB katz} we also show the results for three potentials commonly used for inflation, $\theta_0 \phi^2$, $\theta_0 \phi^4$ and $\theta_{0}\left(1-e^{-\sqrt{2/3}\phi}\right)^2$ (Starobinsky or Higgs inflation). Although within the complexity range of ESR (except Starobinsky for basis set A, where we only reach complexity 8), these potentials have significantly higher description lengths than the best ones we discover, and thus are not highly ranked. For the quadratic and quartic models this is mainly due to the likelihood term, while for Starobinsky or Higgs it is mainly due to the functional complexity, although we do in all cases find functions more accurate than Starobinsky or Higgs. This highlights the power of ESR to uncover novel relations superior to those developed in the literature.

In this analysis, we have been on a quest to find global potentials, i.e. potentials which are defined for all values of $\phi$. There are further restrictions that need to brought in if one is to embed such potentials into a complete theory. One key restriction is that the potential should be bounded from below; if the potential is unbounded then the theory is unstable to vacuum decay and unviable\footnote{Nevertheless, radiative corrections may conspire to stabilize what are, \textit{a priori}, pathological potentials.}.  With this perspective in mind, one can revisit \cref{tab:SetA klogn,tab:SetB klogn,tab:SetA katz,tab:SetB katz} to find that, in the case of the $k \log n$ priors there are only very few functions which violate this criterion, all of which are in Set A. In the case of the Katz prior, again, there are only a few functions which are unbounded. Note also that, for some cases, the potential is only defined for a restricted range of $\phi$; one could envisage this happening in the case where, for example, $\phi$ is the radius of an extra-dimension.

\subsection{Dependence of \texorpdfstring{$r$}{r} on complexity}

\begin{figure*}
    {\includegraphics[width=1\linewidth]{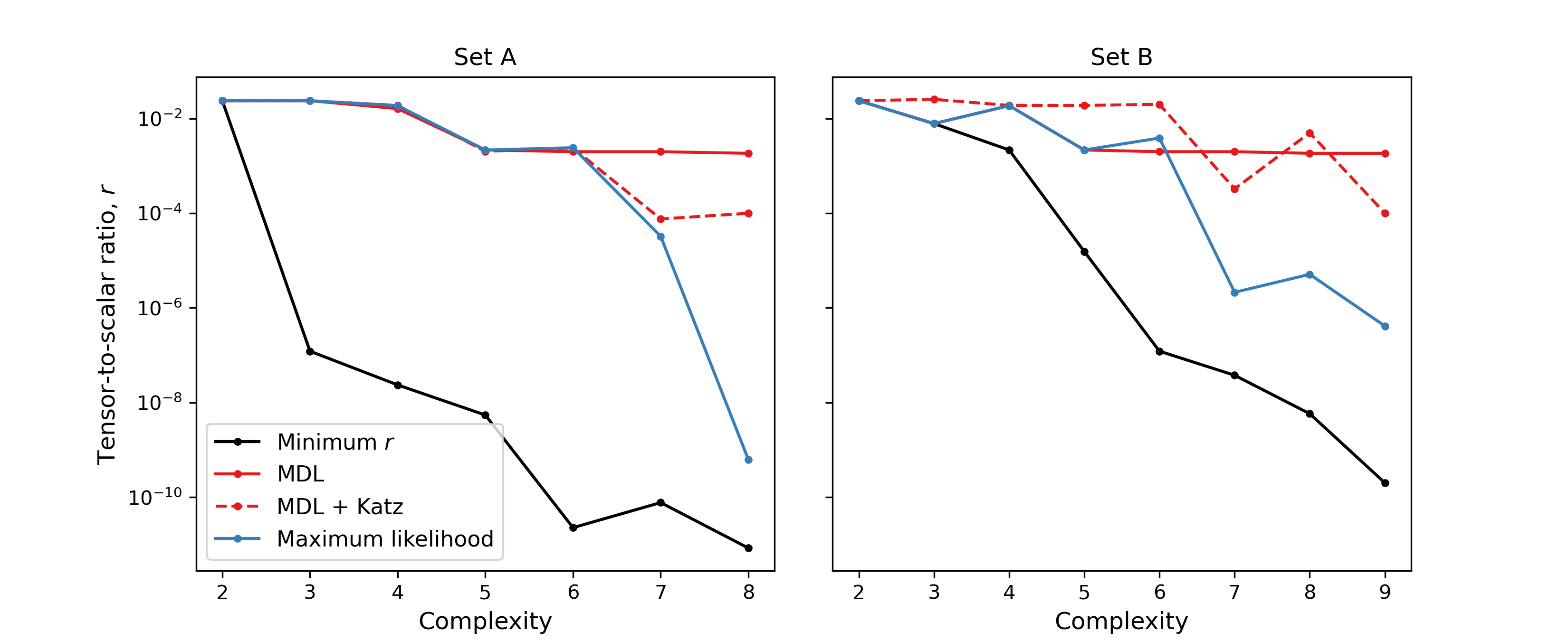}}
  \caption{Variation of the predicted tensor-to-scalar ratio, $r$, with complexity. For both sets of basis operators, the lowest achievable $r$ decreases with the complexity of the potential. We also plot the prediction of the MDL potentials in red, where the result using the $k\log(n)$ prior is solid and with the language model is dashed.
  The blue line shows the predictions of the potentials which maximise the likelihood at a given complexity.
  \label{fig:r_vs_complexity}
  }
\end{figure*}

\begin{figure}
    {\includegraphics[width=1\linewidth]{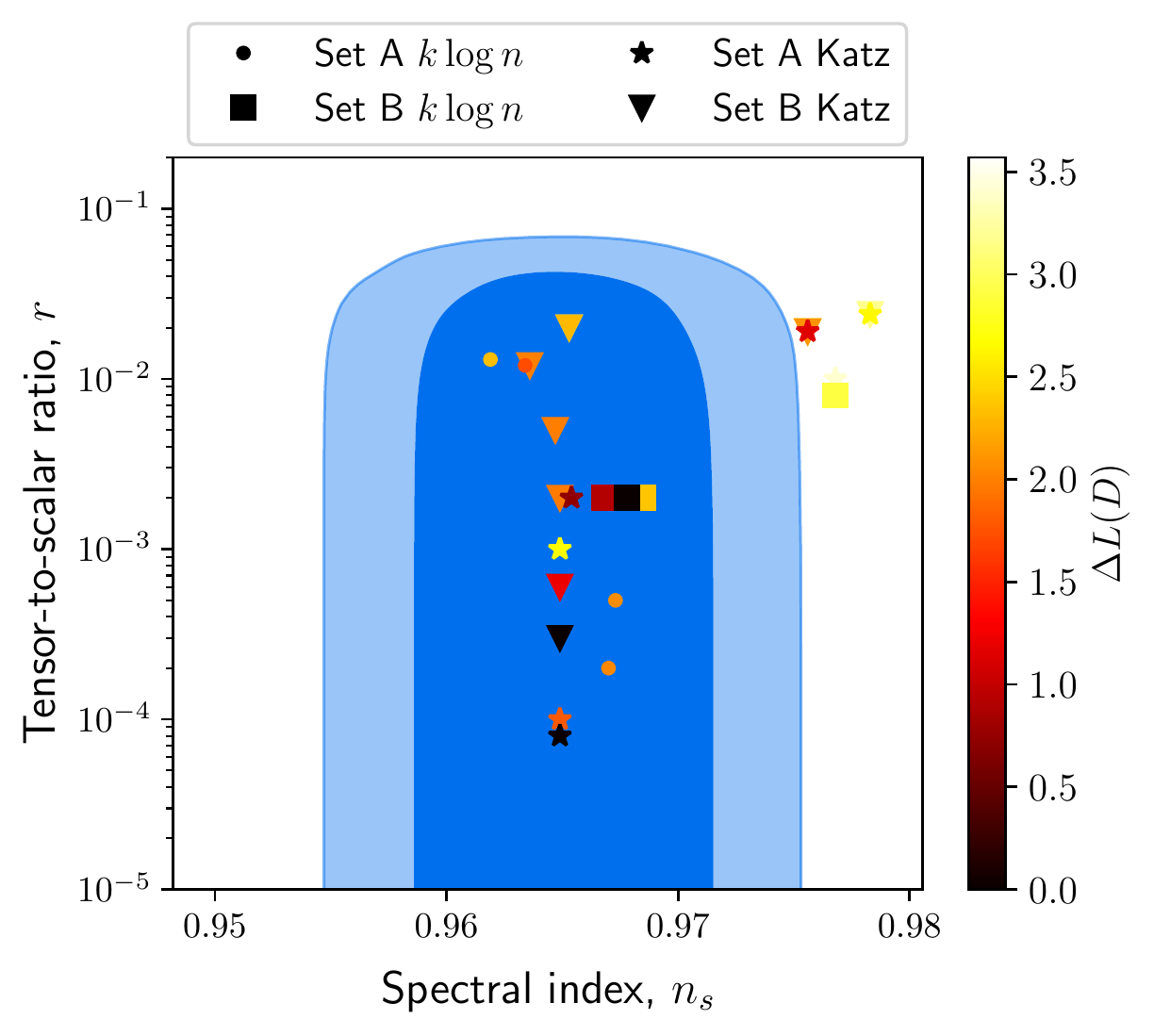}}
    \caption{The predicted tensor-to-scalar ratio and spectral index for the ten best models in each analysis (\cref{tab:SetA klogn,tab:SetB klogn,tab:SetA katz,tab:SetB katz}).
    The 68\% and 95\% CL regions from
    \cref{eq:n_S,eq:r}
    are shown in blue. We colour the points by the change in description length relative to the optimal model for each analysis, such that darker points indicate better potentials.
    }
    \label{fig:planck}
\end{figure}
Let us now revisit the claim in \cite{Boyle:2005ug}, that the {\it phenotypically} simplest models
(e.g. those with the fewest inflection points)
lead to the largest gravitational waves backgrounds, i.e. the largest $r$, by considering whether a similar claim can be made in terms of {\it genotypic} simplicity. A first, general trend to note is that the choice of $k\log n$ versus Katz prior has a mild effect on the value of $r$. In the case of the $k\log n$ prior, the tendency is to favour models with $r\sim 10^{-3}$ (although there are, amongst the top ranked functions, a few cases with higher values of $r$). With the Katz prior, and apart from the highest ranked models, we find more models with $r\sim 10^{-2}$. This indicates that structurally simple models readily allow for very low $r$, while most theoretically motivated models (or at least models that have been proposed in the literature) tend to produce higher values. This is potentially important given that future cosmic microwave background surveys are targeting values within this range.

In \cref{fig:r_vs_complexity} we show how $r$ depends on complexity for both the basis sets. Let us focus first on the the minimum $r$ achievable at a given complexity (solid black line); we do not optimise the functions' parameters specifically to minimise $r$; instead we extract, at each complexity, the function whose full likelihood optimisation yields the lowest value of $r$. We note that for the models lying on this line, we may have values of $n_S$ which are wildly discrepant with the best fit observed values. For this line, we see a clear trend: the larger the complexity, the smaller the $r$. Such a trend is still present if one looks at the models on the Pareto front (the blue line). There, one finds that this dependence is less evident in the case of Set A than in the case of Set B. The dependence of $r$ on complexity is all but washed out once one looks at the Minimum Description Length of models both without and with the language model priors. The minimum value of $r$ can remain unchanged for increasing values of the complexity or even increase for higher complexity.

It is interesting to note that the values of $r$ we are obtaining, either on the Pareto front, or for the Minimum Description Length are already quite low, well within current constraints (as is to be expected) but also close to the expected bounds that will be obtained with future surveys such as from the Simons Observatory \cite{SimonsObservatory:2018koc}. There, one expects to have the ability to place an upper bound such that $r< 3 \times 10^{-3}$ \cite{Simons2019}.
We can already see that, with the low complexities we are probing here, we can already achieve such low values.
In \cref{fig:planck} we illustrate this by plotting the values of ($n_s,r$) for the models in \cref{tab:SetA klogn,tab:SetB klogn,tab:SetA katz,tab:SetB katz} where we can see the values of $r$ plunging to low values.
These models have larger description lengths as they typically need to be more complex to achieve such low values of $r$.

Thus, counter to the claims of \cite{Boyle:2005ug}, it may not be necessary to consider ``unnaturally'' complex potentials to be consistent with the low values of $r$ (see also \cite{Stein:2022cpk}), although it is the case that very low values of $r$ can only be produced by high-complexity models. Finally, we note that, even though a common feature of the potentials we have found here is that they have trans-Planckian evolution of the scalar field ($\Delta\phi>1$), we are still reaching low values of $r$, counter to the conventional wisdom that large $r$ and $\Delta\phi>1$ are indelibly tied together \cite{Lyth:1996im}. These insights demonstrate the advantages offered by an \emph{exhaustive} search through functional parameter space, as afforded only by ESR.

\section{Discussion}
\label{sec:discussion}

\subsection{Caveats and limitations}

An important limitation of our analysis relates to the key concept of complexity.
As discussed in \cref{sec:sr}, the complexity of an expression depends on the choice of basis functions one uses. We have chosen two different sets, with different motivations, to see how the results we obtained might differ. Furthermore, the definition of complexity we use is directly related to the structure of the potential function (its ``genotype''), but one might want to consider alternative definitions that are related to how structured the potential actually is (its ``phenotype''). An alternative approach would therefore be to consider e.g. a linear expansion of the potential in terms of basis functions; the more terms in the expansion, the more structure the function might have, e.g. more inflection points or local minima. In either case (and in general) one cannot be sure that one has reached the global minimum description length, although typically the DL rises toward larger complexity after the minimum.

While the focus of this analysis has been to search for global functions, i.e. potentials, $V(\phi)$ that are valid for all $\phi$, we are well aware that observations only probe a narrow range of the potentials through the few $e$-foldings that affect large scale structure. As we saw in \cref{sec:results}, we can restrict the lists of potentials we found to those that, for example, lead to a positive energy density for all value of $\phi$.

Another physical consideration that could restrict the list further is the dichotomy between ``large'' and ``small'' field models.
While the former models are more robust to initial conditions, the latter are better behaved from an effective field theory point of view \cite{Lyth:1996im}. We have found that, as it currently stands, the selection process of our approach tends to favour large-field models; as argued above, given the low complexity of the models we explore, we tend to evaluate functions of $\phi$ (in Planck units) and not $\theta_i\phi$ where $\theta_i$ is a dimensionless constant that could rescale field variations to become sub-Planckian.

Yet another requirement we have not included is whether the potential contains a natural mechanism for reheating. For example, a potential well at the end of inflation can lead to oscillations in the scalar field trajectory which naturally lend themselves to reheating and pre-heating \cite{Bassett:2005xm}. Potentials incapable of reheating should be disfavoured through an additional criterion.

There is a deeper and more subtle aspect to complexity which must also be addressed. When we look at a theory, the fact that it is simpler may not be immediately apparent from the inspection of its mathematical form in a particular formulation. For example, in the case of Starobinsky inflation \cite{Starobinsky:1980te}, one can either formulate it with scalar fields or in terms of a higher derivative theory of gravity such that the action is
\begin{equation}
    S=\int \sqrt{-g}\:d^4x\:\Big(\frac{M^2_{\rm Pl}}{2}R+\alpha R^2\Big),
\end{equation}
which is, to some extent, less complex.

Another, more notable example of a class of inflationary potentials where the definition of simplicity comes to the fore, is the case of $\alpha$-attractors. There, it has been argued, the models depend on two parameters, but the predictions do not depend on the actual functional form of the potentials \cite{Kallosh:2021mnu}. This means that an approach such as ours, where simplicity is firmly tied to the shape of the potentials, may not capture the complexity of the system.

More generally, there may be physical principles (such as symmetries) that lead to conceptually simple theories with structurally complex potentials.
Nevertheless, our approach has been to search for models that lead to simple expressions as this guiding principle has played an important role in inflationary model building (and physics more generally) in the past. More sophisticated priors along the lines of the Katz model may be used to incorporate any desired theoretical requirements.

\subsection{Possible extensions}

A simple way to address some of the limitations above is to add a prior to the analysis that prefers small-field models and those that include a natural conduit to reheating. This could be done through ``brute force'' by discarding models which do not conform to our requirements, or in a more subtle way through the language model. Specifically, one could restrict the corpus of potentials on which the language models is trained to only small-field models with reheating, and let the prior play its role in reordering the potentials. This would however require a strong correlation between the physical desiderata and the structural properties of the potentials that are picked up by the language model, which is currently unclear.

The approach we have used here, ESR, is restricted to low complexity models.
Given the quality of the current data, and the physical restrictions we are imposing, we are able to explore the optimal models within this range of complexity. But it is conceivable, as one adds more physical constraints, more stringent priors or the data get better (such as with the upcoming Simons Observatory \cite{SimonsObservatory:2018koc}), one may have to explore more complex models. If that is the case one must resort to other methods.

It would already be possible to forecast the results achievable by e.g. the Simon's Observatory by re-running ESR with the constraints it is expected to achieve. Alternatively, one could assume some potential is the correct one, generate data according to it (measured to some precision) and then refit the data with ESR to see what quality of data is necessary to unambiguously pick out the generating function and study the DL differences between the generating function and similar ones (e.g. those with similar $n_s$ and $r$, but simpler). This would provide further information about the likelihood that any of our current best-fitting models are correct, and the data requirements for this to be established conclusively.

If one is to explore more complex models, then a desirable requirement is that the search be {\it exhaustive}, i.e. that one is able to explore {\it all} models of a given complexity before moving to higher complexity. At the moment, the only way which is able to extend the approach of ESR is through grammar enumeration methods \cite[e.g.][]{Kammerer_2021} that systematically explore the space of expressions
by constructing a formal language which can be used to systematically produce all relevant expressions. As opposed to ESR, to reduce the search space the level of nesting of functions is often limited and the grammar designed to avoid duplicate equations which are re-parameterised versions of the same expression (e.g. $\phi \times \theta_0$ and $\phi / \theta_0$). Although one can extend the grammar to allow arbitrary combinations of operators, the latter constraint may prevent one from finding the MDL variant of each expression since changing the parameterisation requires one to specify the parameters to a different precision, thus altering the final terms of \cref{eq:mdl}. We leave such an exploration to future work.

Other methods which are used for symbolic regression, such as Genetic Programming or Reinforcement Learning are more problematic.
For example, as mentioned above, Genetic Programming has a propensity to explore ever more complex expressions, (the ``bloat'' problem) defeating the goal of efficiently seeking simple as well as accurate potentials, and is not guaranteed to find any given good function. There are proposals for reigning in the complexity and directing the searches toward the low-complexity regime but their effectiveness remains to be demonstrated in generality. They may be necessary if inflation is far more complex than considered here.

\section{Conclusion}
\label{sec:conc}

In this paper we have identified
the inflationary potentials that optimally combine simplicity with accuracy given the current data. This is achieved by the \emph{Exhaustive Symbolic Regression} algorithm which systematically generates and evaluates all possible low-complexity expressions given a basis set of operators. We investigated two possible basis sets and two possible priors on them.
This procedure emulates how
an empirico-inductivist
might uncover fundamental physical laws \cite{BayesianMachine}. We find functions that are superior in an information-theoretic sense to other well-known potentials. Further investigation may show these functions to have theoretical significance (and in some cases already have). We have discussed how machine learning-based language models may be used to favour functions similar to those produced in theoretical studies, and anticipate future refinements that will enable to us to home in on functions of particular interest to particle physics model builders.

While we focus in this paper on inflation, the approach may readily be extended to other under-determined problems where the data are not yet sufficiently constraining to single out a particular explanation. This is the situation in many of the open problems in cosmology---most notably the dark matter and dark energy problems---but also true for a wider range of problems in astrophysics, physics, and science.

\section*{Acknowledgements}

We thank Steven Abel, Andrei Constantin, Thomas Harvey, Laura Iacconi, Lukas Kammerer, Gabriel Kronberger,  Andrei Lukas, Adam Moss and David Wands for interesting discussion.

TS was supported by the Oxford Astrophysics Summer Research Programme. DJB is supported by the Simons Collaboration on ``Learning the Universe.''
HD is supported by a Royal Society University Research Fellowship (grant no. 211046).
PGF acknowledges support from STFC and the Beecroft Trust.

For the purpose of open access, the authors have applied a Creative Commons Attribution (CC BY) licence to any Author Accepted Manuscript version arising.

\section*{Data Availability}

The data underlying this article will be shared on reasonable request to the corresponding author.

\bibliography{refs}

\clearpage
\appendix
\begin{widetext}
\section{Tables of ranked functions}
\label{appendix}

\begin{table*}[h!]
\centering
\caption{The ten highest ranked inflationary potentials according to their description lengths for basis set A, using the $k\log(n)$ prior rather than the language model. The predicted $n_s$ and $r$ are at the maximum-likelihood point in each function's parameter space. The rows in red are for expressions which have negative values for some range of $\phi$ or are unbounded;
for functions which are not defined for all $\phi$, we only consider the range of $\phi$ for which these are real. We show also the three components of the DL: the accuracy term (``Residuals''), the structural complexity term (``Function'') and the parametric complexity term (``Parameter''). The ranks apply solely to functions with complexity 1-8 (the range in which we have run ESR), so that entry is left blank for the Starobinsky or Higgs function.
\label{tab:SetA klogn}}
\begin{tabular}{|c|c|c|c|c|c|c|c|c|c|c}
\hline
\multirow{2}{*}{Rank} & \multirow{2}{*}{$V\left( \phi \right) $} & \multirow{2}{*}{Complexity} & \multicolumn{2}{c|}{Prediction} & \multicolumn{4}{c|}{Codelength} \\
\cline{4-9}
\rule{0pt}{3ex}
&  &  & $n_s$ & $r$ & Residuals$^1$ & Function$^2$ & Parameter$^3$ & Total \\
\hline
    1 & $e^{- e^{e^{e^{\phi}}}}$ & 6 & 0.9678 & 0.002 & -12.65 & 6.59 & 0.00 & -6.06 \\
    2 & $\theta_{0} e^{- e^{\phi}}$ & 5 & 0.9668 & 0.002 & -12.79 & 6.93 & 0.69 & -5.16 \\
    3 & $e^{\frac{1}{\sin{\left(\sqrt{\phi} \right)}}}$ & 5 & 0.9634 & 0.012 & -12.39 & 8.05 & 0.00 & -4.35 \\
    4 & $e^{- 3 e^{e^{\phi}}}$ & 6 & 0.9673 & 0.002 & -12.38 & 8.32 & 0.00 & -4.07 \\
    5 & {\color{red} $\theta_{0} - e^{3 \phi}$} & {\color{red}5} & {\color{red}0.9670} & {\color{red}$ 2 \times 10^{-4} $} & {\color{red}-12.77} & {\color{red}8.05} & {\color{red}0.69} & {\color{red}-4.03} \\
    6 & {\color{red}$\theta_{0} - e^{2 \phi}$ }& {\color{red}5} & {\color{red}0.9673 }& {\color{red}$ 5 \times 10^{-4} $} & {\color{red}-12.74} & {\color{red}8.05} & {\color{red}0.69} & {\color{red}-4.00} \\
    7 & $\sin^{2}{\left(\sin{\left(\sin{\left(\sqrt{\phi} \right)} \right)} \right)}$ & 6 & 0.9619 & 0.013 & -12.06 & 8.32 & 0.00 & -3.74 \\
    8 & $e^{e^{\phi} - e^{e^{e^{\phi}}}}$ & 8 & 0.9685 & 0.002 & -12.50 & 8.79 & 0.00 & -3.71 \\
    9 & $\theta_{0} e^{- e^{e^{\phi}}}$ & 6 & 0.9680 & 0.002 & -12.62 & 8.32 & 0.69 & -3.61 \\
    10 & $\left(\theta_{0} + e^{\phi}\right)^{2}$ & 5 & 0.9676 & 0.002 & -12.68 & 8.05 & 1.06 & -3.58 \\
    \vdots & \vdots & \vdots & \vdots & \vdots & \vdots & \vdots & \vdots & \vdots \\
    --- & $\theta_{0}\left(1-e^{-\sqrt{2/3}\phi}\right)^2$ & 9 & 0.9678 & 0.003 & -12.63 & 17.51 & 0.69 & 5.57 \\
    \vdots & \vdots & \vdots & \vdots & \vdots & \vdots & \vdots & \vdots & \vdots \\
    24273 & $\theta_0\phi^2$ & 4 & 0.9669 & 0.132 & 31.82 & 5.55 & 0.69 & 38.06 \\
    \vdots & \vdots & \vdots & \vdots & \vdots & \vdots & \vdots & \vdots & \vdots \\
    38717 & $\theta_0\phi^4$ & 5 & 0.9508 & 0.262 & 168.23 & 6.93 & 0.69 & 175.85 \\

\hline
\end{tabular}
\begin{tabular}{ccc}
     $^1 - \log\mathcal{L} ( \hat{\bm{\theta}} ) $\qquad $^2 k\log(n) + \sum_j \log(c_j)$ &
     \qquad $^3 - \frac{p}{2} \log(3) + \sum_i^p (\log(I_{ii})^{1/2} + \log(|\hat{\theta}_i|))$ \\
\end{tabular}
\end{table*}

\begin{table*}[h!]
\centering
\caption{Same as \cref{tab:SetA klogn} but for basis set B. All of the potentials listed here are non-negative for all $\phi$ for the best-fit parameters.
\label{tab:SetB klogn}}
\begin{tabular}{|c|c|c|c|c|c|c|c|c|c|c}
\hline
\multirow{2}{*}{Rank} & \multirow{2}{*}{$V\left( \phi \right) $} & \multirow{2}{*}{Complexity} & \multicolumn{2}{c|}{Prediction} & \multicolumn{4}{c|}{Codelength} \\
\cline{4-9}
\rule{0pt}{3ex}
&  &  & $n_s$ & $r$ & Residuals & Function & Parameter & Total \\
\hline
    1 & $e^{- e^{e^{e^{\phi}}}}$ & 6 & 0.9678 & 0.002 & -12.65 & 6.59 & 0.00 & -6.06 \\
    2 & $\theta_{0} e^{- e^{\phi}}$ & 5 & 0.9668 & 0.002 & -12.79 & 6.93 & 0.69 & -5.16 \\
    3 & $\left|{\theta_{0}}\right|^{e^{e^{\phi}}}$ & 5 & 0.9674 & 0.002 & -12.72 & 6.93 & 0.69 & -5.09 \\
    4 & $e^{e^{\phi} - e^{e^{e^{\phi}}}}$ & 8 & 0.9685 & 0.002 & -12.50 & 8.79 & 0.00 & -3.71 \\
    5 & $\theta_{0} e^{- e^{e^{\phi}}}$ & 6 & 0.9680 & 0.002 & -12.62 & 8.32 & 0.69 & -3.61 \\
    6 & $e^{- e^{\frac{1}{\phi}}}$ & 5 & 0.9768 & 0.008 & -8.65 & 5.49 & 0.00 & -3.16 \\
    7 & $e^{\theta_{0} e^{e^{\phi}}}$ & 6 & 0.9674 & 0.002 & -12.72 & 8.32 & 1.36 & -3.04 \\
    8 & $\left|{\theta_{0}}\right|^{e^{e^{e^{\phi}}}}$ & 6 & 0.9678 & 0.002 & -12.65 & 8.32 & 1.39 & -2.94 \\
    9 & $e^{e^{\phi}} e^{- e^{e^{e^{\phi}}}}$ & 9 & 0.9685 & 0.002 & -12.50 & 9.89 & 0.00 & -2.62 \\
    10 & $\left|{\theta_{0}}\right|^{\frac{1}{\phi}}$ & 4 & 0.9756 & 0.019 & -8.77 & 5.55 & 0.69 & -2.53 \\
    \vdots & \vdots & \vdots & \vdots & \vdots & \vdots & \vdots & \vdots & \vdots \\
    1272 & $\theta_{0}\left(1-e^{-\sqrt{2/3}\phi}\right)^2$ & 9 & 0.9678 & 0.003 & -12.63 & 17.51 & 0.69 & 5.57 \\
    \vdots & \vdots & \vdots & \vdots & \vdots & \vdots & \vdots & \vdots & \vdots \\
    8697 & $\theta_0\phi^2$ & 4 & 0.9669 & 0.132 & 31.82 & 5.55 & 0.69 & 38.01 \\
    \vdots & \vdots & \vdots & \vdots & \vdots & \vdots & \vdots & \vdots & \vdots \\
    10839 & $\theta_0\phi^4$ & 5 & 0.9508 & 0.262 & 168.23 & 6.93 & 0.69 & 178.81 \\
\hline
\end{tabular}
\end{table*}

\begin{table*}[h!]
\centering
\caption{The ten best inflationary potentials according to their description lengths for basis set A after applying a language model prior on the functions. Again, we colour potentials red if they have negative values for the maximum likelihood parameters or are unbounded, and the column headings are defined in \cref{tab:SetA klogn}. In principle the first ranked potential could be positive for all $\phi$, however the maximum likelihood value of $\theta_0$ is negative, such that $V(\phi)\to -\infty$ as $\phi \to \infty$.
\label{tab:SetA katz}}
\begin{tabular}{|c|c|c|c|c|c|c|c|c|c|c}
\hline
\multirow{2}{*}{Rank} & \multirow{2}{*}{$V\left( \phi \right) $} & \multirow{2}{*}{Complexity} & \multicolumn{2}{c|}{Prediction} & \multicolumn{4}{c|}{Codelength} \\
\cline{4-9}
\rule{0pt}{3ex}
&  &  & $n_s$ & $r$ & Residuals & Function & Parameter & Total \\
\hline
    1 & {\color{red}$\theta_{0} \left(\theta_{1} + \log{\left(\phi \right)}^{2}\right)$} & {\color{red}7 }& {\color{red}0.9649 }& {\color{red}$ 8 \times 10^{-5} $ }& {\color{red}-12.90} & {\color{red}9.28} & {\color{red}1.39} & {\color{red}-2.23} \\
    2 & $\frac{1}{\left(\theta_{0} + e^{\phi}\right)^{2}}$ & 6 & 0.9654 & 0.002 & -12.88 & 10.31 & 1.06 & -1.51 \\
    3 & $e^{\frac{\theta_{0}}{\phi}}$ & 4 & 0.9756 & 0.019 & -8.77 & 6.98 & 0.69 & -1.10 \\
    4 & $\theta_{0} \left(\theta_{1} - e^{\phi}\right)^{2}$ & 7 & 0.9676 & 0.002 & -12.68 & 10.77 & 1.39 & -0.53 \\
    5 & $\frac{\theta_{0}}{\phi + \log{\left(\frac{\left|{\theta_{1}}\right|}{\phi} \right)}}$ & 8 & 0.9649 & $ 1 \times 10^{-4} $ & -12.90 & 11.06 & 1.39 & -0.45 \\
    6 & $\left(\theta_{0} - e^{\phi}\right)^{2}$ & 5 & 0.9676 & 0.002 & -12.68 & 11.59 & 1.06 & -0.04 \\
    7 & {\color{red}$\theta_{0} \log{\left(\phi \right)}$} & {\color{red}4} & {\color{red}0.9783} & {\color{red}0.024} & {\color{red}-6.39} & {\color{red}6.10} & {\color{red}0.69} & {\color{red}0.40} \\
    8 & $\frac{1}{\theta_{0} + e^{\frac{\phi}{\theta_{1}}}}$ & 7 & 0.9649 & 0.001 & -12.90 & 11.23 & 2.08 & 0.42 \\
    9 & $\theta_{0} - \frac{\theta_{1}}{\phi}$ & 5 & 0.9768 & 0.010 & -8.62 & 8.12 & 1.65 & 1.16 \\
    10 & $\theta_{0} e^{\theta_{1} e^{\phi}}$ & 7 & 0.9668 & 0.002 & -12.79 & 12.74 & 1.39 & 1.34 \\
    \vdots & \vdots & \vdots & \vdots & \vdots & \vdots & \vdots & \vdots & \vdots \\
    16653 & $\theta_0\phi^2$ & 4 & 0.9669 & 0.132 & 31.82 & 5.49 & 0.69 & 38.01 \\
    \vdots & \vdots & \vdots & \vdots & \vdots & \vdots & \vdots & \vdots & \vdots \\
    37635 & $\theta_0\phi^4$ & 5 & 0.9508 & 0.262 & 168.23 & 7.15 & 0.69 & 176.08 \\
    \vdots & \vdots & \vdots & \vdots & \vdots & \vdots & \vdots & \vdots & \vdots \\
    --- & $\theta_{0}\left(1-e^{-\sqrt{2/3}\phi}\right)^2$ & 9 & 0.9678 & 0.003 & -12.63 & 11.42 & 0.69 & -0.52 \\

\hline
\end{tabular}

\end{table*}

\begin{table*}[h!]
\centering
\caption{Same as \cref{tab:SetA katz} but for basis set B. \label{tab:SetB katz}}
\begin{tabular}{|c|c|c|c|c|c|c|c|c|c}
\hline
\multirow{2}{*}{Rank} & \multirow{2}{*}{$V\left( \phi \right) $} & \multirow{2}{*}{Complexity} & \multicolumn{2}{c|}{Prediction} & \multicolumn{4}{c|}{Codelength} \\
\cline{4-9}
\rule{0pt}{3ex}
& & & $n_s$ & $r$ & Residuals & Function & Parameter & Total \\
\hline
    1 & $\theta_{0} \phi^{\frac{\theta_{1}}{\phi}}$ & 7 & 0.9649 & $ 3 \times 10^{-4} $ & -12.90 & 8.92 & 1.39 & -2.59 \\
    2 & $\theta_{0} \left(\theta_{1} + \phi^{\phi}\right)$ & 7 & 0.9649 & $ 6 \times 10^{-4} $ & -12.90 & 10.12 & 1.39 & -1.39 \\
    3 & $\theta_{0} \phi^{\phi^{\theta_{1}}}$ & 7 & 0.9649 & 0.002 & -12.89 & 9.55 & 2.71 & -0.63 \\
    4 & $\phi^{\theta_{0}} e^{\frac{\theta_{1}}{\phi}}$ & 8 & 0.9647 & 0.005 & -12.83 & 9.37 & 2.83 & -0.62 \\
    5 & $\left(\phi + e^{\frac{\theta_{0}}{\phi}}\right)^{\theta_{1}}$ & 8 & 0.9636 & 0.012 & -12.48 & 10.05 & 1.83 & -0.60 \\
    6 & $\left|{\theta_{0}}\right|^{\frac{\theta_{1}}{\phi}}$ & 5 & 0.9756 & 0.019 & -8.77 & 6.88 & 1.39 & -0.50 \\
    7 & $e^{\theta_{0} \left|{\theta_{1}}\right|^{\phi}}$ & 8 & 0.9653 & 0.020 & -11.66 & 8.11 & 3.26 & -0.30 \\
    8 & $e^{\frac{\theta_{0}}{\phi}}$ & 4 & 0.9756 & 0.019 & -8.77 & 8.04 & 0.69 & -0.03 \\
    9 & $\theta_{0} e^{\frac{\theta_{1}}{\phi}}$ & 6 & 0.9653 & 0.020 & -8.82 & 7.84 & 1.39 & 0.40 \\
    10 & $\theta_{0} \log{\left(\phi \right)}$ & 4 & 0.9783 & 0.024 & -6.39 & 6.26 & 0.69 & 0.57 \\
    \vdots & \vdots & \vdots & \vdots & \vdots & \vdots & \vdots & \vdots & \vdots \\
    12 & $\theta_{0}\left(1-e^{-\sqrt{2/3}\phi}\right)^2$ & 9 & 0.9678 & 0.003 & -12.63 & 12.64 & 0.69 & 0.70 \\
    \vdots & \vdots & \vdots & \vdots & \vdots & \vdots & \vdots & \vdots & \vdots \\
    5401 & $\theta_0\phi^2$ & 4 & 0.9669 & 0.132 & 31.82 & 4.67 & 0.69 & 38.03 \\
    \vdots & \vdots & \vdots & \vdots & \vdots & \vdots & \vdots & \vdots & \vdots \\
    8938 & $\theta_0\phi^4$ & 5 & 0.9508 & 0.262 & 168.23 & 4.67 & 0.69 & 180.84 \\
\hline
\end{tabular}
\end{table*}

\end{widetext}
\end{document}